\pdfoutput=1
\documentclass{emulateapj}
\usepackage[varg]{txfonts}
\usepackage{natbib}
\usepackage{microtype}
\DeclareRobustCommand\New[1]{#1}
\DeclareRobustCommand\NewNew[1]{#1}
\begin{document}

\title{\New{Can} solid body destruction explain abundance discrepancies 
  in planetary nebulae\New{?}}
\author{William J. Henney\altaffilmark{1} and Gra\.zyna Stasi\'nska\altaffilmark{2}}
\altaffiltext{1}{Centro de Radioastronom\'{\i}a y Astrof\'{\i}sica,
  Univer\-si\-dad Na\-cio\-nal Aut\'o\-no\-ma de M\'exi\-co, Morelia,
  M\'exico}
\altaffiltext{2}{LUTH, Observatoire de Paris, CNRS, Universit\'e Paris Diderot;
     Place Jules Janssen 92190 Meudon, France}

\shorttitle{SOLID BODY DESTRUCTION IN PLANETARY NEBULAE}
\shortauthors{HENNEY \& STASI\'NSKA}
\email{w.henney@crya.unam.mx, grazyna.stasinska@obspm.fr}

\begin{abstract}
  In planetary nebulae, abundances of oxygen and other heavy elements derived from optical recombination lines are systematically higher than those derived from collisionally excited lines.  We investigate the hypothesis that the destruction of solid bodies may produce pockets of cool, high-metallicity gas that could explain  these abundance discrepancies. Under the assumption of maximally efficient radiative ablation, we derive two fundamental constraints that the solid bodies must satisfy in order \New{that their evaporation during the planetary nebula phase should} generate a high enough gas phase metallicity. A local constraint implies that the bodies must be larger than tens of meters, while a global constraint implies that the total mass of the solid body reservoir must exceed a few hundredths of a solar mass. This mass greatly exceeds the mass of any population of comets or large debris particles expected to be found orbiting evolved low- to intermediate-mass stars. We therefore conclude that \New{contemporaneous} solid body destruction cannot explain the observed abundance discrepancies in planetary nebulae. \New{\NewNew{However,} similar arguments 
\NewNew{applied to} the sublimation of solid bodies during the preceding asymptotic giant branch (AGB) phase \NewNew{do not lead to such a clear-cut conclusion}. \NewNew{In this case, the required reservoir of volatile solids is only one ten-thousandth of a solar mass, which is comparable to the most massive debris disks observed around solar-type stars, implying that this mechanism may contribute to abundance discrepancies in at least some planetary nebulae, so long as mixing of the high metallicity gas is inefficient.} 
}

\end{abstract}
\maketitle

\section{Introduction}
\label{sec:introduction}

In recent years, evidence has mounted for very large differences between the heavy-element abundances calculated in planetary nebulae (PNe) from optical recombination lines (ORLs) and those calculated from collisionaly excited lines (CELs) for  C, N, O and Ne ions. Until the origin of these differences is completely understood, there will be a doubt about the accuracy of abundances derived from nebular spectra. The most comprehensive and accurate set of observational data has been obtained for the O\(^{2+}\) ion, which emits both \ion{O}{2} ORLs and [\ion{O}{3}] CELs in the visual. The ratio between the O\(^{2+}\)/H\(^+\) abundance derived from ORLs and CEls is termed the \textit{abundance discrepancy factor} (ADF):
\[
\mathrm{ADF} = \frac{
 ( \mathrm{O^{2+}} / \mathrm{H^+} )_\mathrm{ORL}
 }{
 ( \mathrm{O^{2+}} / \mathrm{H^+} )_\mathrm{CEL}
 }
\]
The present status of our knowledge of ADFs is summarized in \citet{2005ApJS..157..371R} and \citet{2002RMxAC..12...70L, 2006IAUS..234..219L, 2006pnbm.conf..169L}. ADFs for the O\(^{2+}\) ion have been measured in about 90 PNe. Their distribution peaks at a value of 2--3, and shows a long tail of values exceeding 10. The ADFs tend to be larger for more evolved nebulae, and, in the few cases where they have been determined at different positions in the same nebula, they are larger closer to the central star.
\New{ADFs have also been found in HII regions, although never larger than about a factor of 2 \citep{2007ApJ...670..457G}.  There is so far no fully convincing explanation for these ADFs, and it is not certain they arise from the same physical causes in the two families of objects. It has recently been suggested that irradiation by X-rays in photoionzed nebulae could produce ADFs, but the available X-ray luminosity is too small by a large factor \citep{Ercolano:2009}.} 
\New{Another} possibility is that the recombination lines arise mainly from cool, H-poor knots in the ionized gas. The chemical composition of those knots rules out that they could be made of nucleo-processes material ejected by the PN progenitor \citep{2000MNRAS.312..585L, 2005MNRAS.362..424W}. One attractive suggestion \citep{2003IAUS..209..339L, 2006pnbm.conf..169L} is that they are produced by the evaporation of solid bodies (asteroids, comets, planets) belonging to the planetary system of the PN progenitor. In this paper, we examine this suggestion quantitatively, by establishing the conditions necessary for such an hypothesis to work. We concentrate on oxygen as the element for which the most reliable observational evidence exists. We first (\S~\ref{sec:formal-constraints}) establish purely formal constraints on the size and total mass of the solid bodies in order that their ablation might produce a cool, high-metallicity gaseous phase able to reproduce the observed ADFs. We then (\S~\ref{sec:appl-plan-nebul}) consider in detail the physical processes reponsible for solid body ablation in PNe, estimating the magnitude of various terms that enter into the constraints of \S~\ref{sec:formal-constraints}. In \S~\ref{sec:appl-minim-size} we consolidate the previous sections to derive quantitative lower limits on the solid body size and total reservoir mass. The implications of our results are discussed in \S~\ref{sec:discussion}. For convenience, a list of symbols used in the paper is given in the appendix.

\section{Formal constraints}
\label{sec:formal-constraints}

\subsection{How large must the solid bodies be?}
\label{sec:length-scales}
A continuum fluid description of a collisional plasma is only valid for scales much larger than the mean free path for interactions between the particles, \(\ell\). A distinct thermodynamic temperature can also be defined only at scales \(\gg \ell\), since multiple collisions are required in order for the kinetic energies of the particles to reach equilibrium. Quantitative estimates of these length scales for nebular conditions are given in \S~\ref{sec:part-inter-scale}.

\newcommand\ab{_\mathrm{ab}} 
\newcommand\Ox{_\mathrm{O}} 
\newcommand\neb{_1} 
\newcommand\sol{_\mathrm{solid}} 
\newcommand\ii{_\mathrm{i}}
\newcommand\orl{_\mathrm{\scriptscriptstyle ORL}} 

Suppose that some ablation process acts on a solid body of radius \(a\) so as to produce an ablative flux, \(F\ab\), of metal atoms, which leave the solid surface at velocity \(v\ab\). In order to produce a distinct high-metallicity, low-temperature phase, it is necessary for the ablated atoms to have a high concentration on scales \(> \ell\). On scales smaller than \(\ell\), the ablated atoms will stream freely,\footnote{Complications arising from the coupling of charged particles to the magnetic field are considered in \S~\ref{sec:part-inter-scale}.} so their number density will be given by \(n\ab = F\ab a^2 / v\ab r^2\), where \(r\) is the distance from the solid body. If the hydrogen number density in the nebula is \(n\), then the mean metallicity enhancement by number over the volume \(\ell^3\) is given by 
\begin{equation}
  z\ab(\ell) 
  \equiv \left\langle \frac{n\ab}{n} \right\rangle_{\ell^3}
  = \frac{F\ab}{v\ab n} \frac{3 a^2}{\ell^2} 
.\label{eq:zlocal}
\end{equation}
In this expression, the term \( F\ab/(v\ab n)\) is the gas-phase metallicity enhancement at the surface of the solid body, while the term \(3 a^2/\ell^2\) represents the dilution over the macroscopic volume \(\ell^3\), assuming \(\ell > a\). If \(\ell \le a\) then this term would be unity.\footnote{Equation~(\ref{eq:zlocal}) also
  assumes that \(s > \ell\), where \(s\) is the mean separation
  between solid bodies. If this is not the case, then the right-hand
  side must be multiplied by \(\ell^3/s^3\) due to crowding effects.\label{fn:crowding}} The condition that the metal concentration must be sufficiently high on macroscopic scales can be expressed as \(z\ab(\ell) \gg z\), where \( z\) is the ``background'' metallicity of the nebula, implying that 
\begin{equation}
  \label{eq:a-over-l}
  a \gg \left( \frac{z n v\ab}{3 F\ab}\right)^{1/2} \ell . 
\end{equation}
This criterion establishes a minimum radius that an ablating solid body must satisfy in order to produce macroscopic ``pockets'' of metal-enriched gas of metallicity \(\gg z\). In \S~\ref{sec:appl-plan-nebul} numerical values of this radius will be estimated for planetary nebula conditions.

\subsection{How much oxygen must be in the cool phase?}
\label{sec:minimum-oxygen-mass}

In this section, we estimate what is the minimum oxygen mass that must be present in the cool, high metallicity gas in order to produce a significant ADF\@. This is independent of the origin of the high metallicity phase.

Consider an idealized two-phase model for an ionized nebula, comprising the normal nebular gas (phase 1) plus a phase that is enriched in metals (phase 2). The metallicity, density, and temperature of each phase will be different, but we assume that they are in thermal pressure equilibrium. To be definite, we characterize the metallicity by \(z\), the oxygen abundance by number relative to hydrogen. So long as \(z \ll 1\) in both phases and hydrogen is fully ionised, both the electron density and the total ion density will be roughly equal to the hydrogen density, \(n\).
In this approximation, thermal pressure balance implies \begin{equation} 
  n_1 T_1 \simeq n_2 T_2 .\label{eq:press-bal} 
\end{equation}

The total oxygen mass of phase \(i\) (\(i = 1\) or \(2\)) is given by \begin{equation} M_{\mathrm{O},i} = m n_i z_i V_i ,\label{eq:ox-mass} \end{equation} where \(V_i\) is the total volume occupied by phase \(i\) and \(m = 16 m_\mathrm{H}\) is the mass of an oxygen atom. The total luminosity (photons~s\(^{-1}\)) from phase \(i\) of a given \ion{O}{2} optical recombination line is 
\begin{equation} 
  L_i \simeq n_i^2 y_i z_i \, \alpha\orl (T_i) \, V_i,\label{eq:orl-lum}
\end{equation} 
where \(y_i\) is the ionization fraction of O\(^{2+}\) in phase \(i\) and \(\alpha\orl\) is the appropriate effective recombination coefficient, which is assumed to depend on temperature as 
\begin{equation}
  \alpha\orl (T) \simeq \alpha_0 (T/T_0)^{-b} \label{eq:alpha-vs-T} , 
\end{equation}
where \(\alpha_0\), \(T_0\), and \(b \simeq 1\) are constants.

A necessary condition for producing a significant ADF is that the total optical recombination line luminosity of phase 2 should be at least equal to that of phase~1: \(L_2 \ge L_1\). In the limit that most of the emission of collisionally excited metal lines and hydrogen recombination lines comes from phase 1, this would give an ADF of at least two. Using equation~(\ref{eq:orl-lum}), this condition can be expressed as \begin{equation}
  \label{eq:lum-condition}
  \frac{n_2^2 y_2 z_2 \, \alpha\orl (T_2)  \, V_2}
  {n_1^2 y_1 z_1 \, \alpha\orl (T_1) \, V_1}
  \ge 1
\end{equation}
or, substituting equations~(\ref{eq:press-bal}), (\ref{eq:ox-mass}),
and (\ref{eq:alpha-vs-T}),  
\begin{equation}
  \label{eq:mass-condition}
  \frac{M_{\mathrm{O},2}}{M_{\mathrm{O},1}} \ge \mu_{\mathrm{min}}
  \quad \mathrm{where} \quad 
  \mu_\mathrm{min}  = \frac{y_1}{y_2} \left(\frac{T_2}{T_1}\right)^2
  .
\end{equation}
Numerical estimates for this minimum mass fraction, \(\mu_{\mathrm{min}}\) will be provided in \S~\ref{sec:appl-plan-nebul}. 



\subsection{How much mass must be in the solid body reservoir?}
\label{sec:reservoir-mass}

The timescale for the total destruction via ablation of a solid body of radius \(a\) and density \(\rho\sol\) will be
\begin{equation}
t\ab \equiv \frac{m\sol}{d m\sol / dt} 
= \frac{\frac43 \pi a^3 \rho\sol}{\pi a^2 F\ab m\ab} 
= \frac{4 a \rho\sol}{3 F\ab m\ab}, \label{eq:ablation-timescale}
\end{equation}
where \(m\ab\) is the mass of each ablated atom. Thus, during the lifetime of the nebula, \(t\), the total mass of ablated metals will be 
\begin{equation}
M\ab \simeq  \min(1, t/t\ab) M\sol, \label{eq:ablated-vs-reservoir}
\end{equation}
where \(M\sol\) is the initial total mass of the reservoir of solid bodies.


The issue of the mixing of the ablated metals into photoionized gas is a complicated one \citep[e.g.,][]{2007A&A...471..193S}, but a lower limit on the required mass reservoir can be obtained by simply assuming that no mixing occurs, perhaps because the mixing timescale is greater than the nebular age. On that assumption, and if  \(f\Ox\) is the oxygen mass  fraction of the solid bodies, then \(f\Ox M\ab\) can be identified with the minimum mass of oxygen, \(M_{\mathrm{O},2}\), which must be present in the cool gas phase in order to produce a significant ADF (see \S~\ref{sec:minimum-oxygen-mass}). At the same time, the oxygen mass, \(M_{\mathrm{O},1}\) in the background nebula is \(M_{\mathrm{O},1} \simeq 16 z\neb M\neb\), so that the gas-phase minimum mass fraction condition of equation~(\ref{eq:mass-condition}), when combined with equation~(\ref{eq:ablated-vs-reservoir}), gives the following minimum mass of the solid body reservoir:
\begin{equation}
  \label{eq:ablated-minimum-mass}
   M\sol \ge 16 \, (t\ab/t) \, z\neb \mu_{\mathrm{min}} M\neb / f\Ox  . 
\end{equation}
Note that if the solid body ablation timescale greatly exceeds the nebular age (\(t\ab \gg t\)), then the mass required to be locked up in the solid body reservoir will be much larger than the gas-phase oxygen mass  required to explain the ADF (\S~\ref{sec:minimum-oxygen-mass}). 

\renewcommand\neb{} 

\section{Physical constraints}
\label{sec:appl-plan-nebul}

Now that we have established the formal constraints on the solid body size, eq.~(\ref{eq:a-over-l}), and total reservoir mass, eq.~(\ref{eq:ablated-minimum-mass}), that must be satisfied in order for solid body destruction to explain the observed ADFs, we are in a position to calculate quantitative values appropriate to the physical conditions in PNe. To that end, we first estimate values for various parameters that enter in these two equations: the minimum gas-phase oxygen fraction, \( \mu_{\mathrm{min}}\); the interaction length-scale \(\ell\); and the flux and velocity of ablated particles, \(F\ab\) and \(v\ab\). In anticipation of our final results, we are most interested in finding robust limits on these parameters such that the effects of solid body destruction are maximised. 

In this and following sections, we frequently make estimates based on typical physical conditions in PNe. To show how the results vary with the nebular parameters, we employ dimensionless scaled variables as follows: ionized gas density \(n = 10^4 n_4~\mathrm{cm^{-3}}\); ionized gas temperature \(T = 10^4 T_4~\mathrm{K}\); central star ionizing luminosity \(S_* = 10^{47} S_{47}~\mathrm{s^{-1}}\); central star bolometric luminosity \(10^4 L_4~L_\odot\) ; radius from central star \(R = 10^{17} R_{\mathrm{17}}~\mathrm{cm}\); nebular expansion velocity \(v = 20 v_{20}~\mathrm{km\ s^{-1}}\).

\subsection{Minimum gas-phase oxygen fraction, \(\mu_{\mathrm{min}}\)}
\label{sec:minimum-mass-needs}

Photoionization models indicate\New{\footnote{\New{Calculations using two different photoionization codes \citep{1998PASP..110..761F, 2005A&A...434..507S} give consistent results.}}} that typical values of \(T_2/T_1 \le 0.1\) are achieved if \(z_2\) is enriched by a factor of 100 to 1000 with respect to \(z_1\), which is assumed to be roughly at the solar value. A lower limit for \(y_1/y_2\) can be estimated under the assumption that the abundance of O\(^{+}\) is negligible in both zones,\footnote{A reasonable assumption for the hard ionizing spectra typical of PNe.} so that the oxygen ion balance is solely between O\(^{2+}\) and O\(^{3+}\). From equation~(\ref{eq:mass-condition}), it can be seen that the smallest values of \(\mu_{\mathrm{min}}\) will be obtained when there is little O\(^{3+}\) in the cool phase~2, so that \(y_2 \sim 1\). To produce an appreciable ADF, it is necessary for the collisional O\(^{2+}\) lines to come from phase~1, so that \(y_1\) cannot be exceedingly low. We therefore take a conservative lower limit of \(y_1/y_2 \ge 0.1\), which, when combined with an estimated \(T_2 / T_1 \sim 0.1\), gives a lower limit to the oxygen mass fraction of \(\mu_\mathrm{min} > 0.001\).

\subsection{Particle interaction scale, \(\ell\)}
\label{sec:part-inter-scale}

For the charged particles in a photoionized plasma, the interparticle coupling is dominated by the accumulation of many small-angle deviations, mediated by the long range Coulomb force. These electrostatic interactions are effectively shielded at separations larger than the Debye length, giving an effective collisional mean free path of \(\ell_\mathrm{C} \simeq 6.1 \times 10^7 T_4^2 / n_4 ~\mathrm{cm}\), where the weak dependence on physical conditions of the Coulomb logarithm has been suppressed. In addition, in the presence of magnetic fields, charged particles are strongly coupled to the field lines on the scale of the gyroradius:  \(\ell_\mathrm{L} = 1.25 \times 10^5 (\beta A\ab/n_4)^{1/2} ~\mathrm{cm}\), where \(\beta\) is the ratio of thermal to magnetic pressure, which is likely to be \(\sim 1\) but with strong local variations. If the magnetic fields are turbulent, then there may be indirect particle-particle coupling on scales \(\sim \ell_\mathrm{L}\). 

\subsection{Particle ablation rate, \(F\ab\)}
\label{sec:maxim-ablat-rate}

The mechanisms for the ablation of volatiles from solid bodies in a planetary nebula may include thermal sublimation (evaporation), sputtering by particle collisions, and photosputtering (photodesorption) by ultraviolet photons. To be definite, we will consider two canonical cases for the nature of the ablating particles: (1) the case of water ice, where a relatively small amount of energy is required to break molecular bonds and liberate particles from the surface \citep{1995P&SS...43.1311W}, and (2) the case of amorphous silicates, where much more energetic events are required in order to detach particles from the solid \citep{Field:1997}. These will be referred to as the \textit{volatile} case and \textit{refractory} case, respectively.


\subsubsection{EUV photosputtering and the maximal ablation rate} 
\label{sec:maxim-phot-rate}

Photosputtering, also termed photodesorption in the case of volatiles, is the ejection of heavy particles (e.g., ions or molecules) due to the interaction of individual photons with a solid surface.  The flux of ablated metals, \(F\ab\), will be equal to the incident flux of photons in the relevant spectral range multiplied by \(Q_{\mathrm{abs}} Y\ab\), where \(Q_{\mathrm{abs}}\) is the fraction of photons that are absorbed by the surface and \(Y\ab\) is the yield. Solid bodies that are much larger than the photon wavelength generally have a low albedo, so that \(Q_{\mathrm{abs}} \sim 1\), but the yield is a strong function of the incident photon energy, being essentially zero for photon energies below a threshold that is of order the solid binding energy. The photosputtering process is similar in many ways to the photoelectric effect, which has been much more extensively studied, with the difference that it is ions instead of electrons that are ejected in photosputtering. This difference is crucial because the electrons strongly couple to the electromagnetic field of the photons, leading to a photoelectric yield of order unity for photons with energies above the electron binding energy of \(\sim 10\)~eV. The photosputtering yield is much more uncertain since no relevant laboratory or theoretical studies exist. It is likely to be significantly smaller than unity, at least in the case of refractory material, since the relatively poor coupling between solid ions and the photons makes it difficult for them to be ejected. However, a firm upper limit on the ablation rate can be obtained by setting \(Y\ab = 1\), which may be a realistic value for volatile materials (where photochemical reactions can enhance the coupling) but is probably an overestimate for refractory solids.

For the hot central stars of PNe, the majority of the emitted photons are above the hydrogen-ionizing limit of 13.6~eV, so that the maximal ablated flux is simply equal to the ionizing stellar flux:
\begin{equation}
  \label{eq:maximal-ablated-flux}
  F\ab < F_* = \left(S_* / 4 \pi R^2\right) e^{-\tau}  = 7.96 \times 10^{11} S_{47} R_{17}^{-2} e^{-\tau}~\mathrm{cm^{-2}\ s^{-1}}
\end{equation}
where \(\tau\) is the neutral hydrogen optical depth, which is small except for near the edge of an ionization-bounded nebula.

\subsubsection{Contribution of other ablation mechanisms}
\label{sec:contr-therm-evap}
A fundamental assumption of the above analysis is that the ablation occurs via a single-particle photon process and is limited by the stellar radiation flux impinging on the solid body. However, it is also necessary to consider other ablation processes, such as particle sputtering and thermal sublimation, as well as the contribution of the diffuse radiation field to the UV flux. 

\paragraph{Lyman \(\alpha\) photosputtering}

The diffuse flux of ionizing EUV radiation is likely to be only a small fraction of the direct stellar flux \citep{2001ApJ...548..288L} and will not contribute significantly to the total photosputtering rate. However, non-ionizing far ultraviolet (FUV) radiation can also contribute to sputtering, especially for volatile materials such as ices, and this may be dominated by the diffuse component. As an example, we calculate the diffuse flux of the Lyman \(\alpha\) line at 912~\AA{}, with a photon energy of \(10.2\)~eV\@.

At typical nebular temperatures, recombination greatly exceeds collisional excitation as a source for Ly\(\alpha\), whereas, for Galactic metallicity, the principal sink is dust absorption rather than escape in the line wings \citep{1980ApJ...236..609H, 1998AJ....116..322H, 2007dmsf.book..103H}. Assuming a balance between sources and sinks, the diffuse Ly\(\alpha\) flux on an opaque obstacle is found to be 
\begin{equation}
  \label{eq:lya-flux}
  F_\alpha \simeq n \alpha_{\mathrm{B}} / 4 \sigma_{\mathrm{d}} , 
\end{equation}
where \(\alpha_{\mathrm{B}}\) is the hydrogen recombination coefficient to excited levels, and \( \sigma_{\mathrm{d}} \simeq 10^{-21}~\mathrm{cm^2\ H^{-1}}\) is the dust absorption cross section per hydrogen nucleon in the FUV\@. The area that the solid body presents to the diffuse radiation is \(4 \pi a^2\), as opposed to \(\pi a^2\) presented to the stellar radiation, so that the ratio of the rate of inicidence of diffuse to stellar photons is \(4 F_\alpha / F_* \). This is inversely proportional to the ionization parameter, defined as \(U = {F_*}/{c n\neb}\). The value of \(U\) will vary with radius inside the nebula, but for an ionization-bounded, uniform density nebula, the mean value will be \(U = 6.2 \times 10^{-3} \left(S_{47} / R_{17}  T_4\right)^{1/2} \). Therefore, we find 
\begin{equation}
  \label{eq:lya-ratio}
  4 F_\alpha / F_* = 0.0085 (T_4 \sigma_{-21} U)^{-1} = 1.35 R_{17}^{1/2} S_{47}^{-1/2} T_4^{-1/2} \sigma_{-21}^{-1}, 
\end{equation}
from which it can be seen that \(4 F_\alpha\)  typically exceeds \(F_*\), but by less than a factor of ten. However, in order to translate from \(F_\alpha\) to an ablated particle flux \(F_{\mathrm{ab}}\) it is necessary to multiply by the yield, which in the case of FUV photons is well-determined experimentally and found to be \(Y_{\mathrm{ab}} = 10^{-3}\) to \(10^{-2}\) for water ice \citep{1995P&SS...43.1311W}. This is much lower than the \(Y_{\mathrm{ab}} = 1\) that was assumed when calculating the maximal photosputtering flux,\footnote{It is entirely possible that the sputtering yield for ionizing EUV photons is also \(\ll 1\), in which case the total ablation flux would have to be revised downwards from its maximal value. However, we know of no relevant experimental studies that address this question.} so that inclusion of sputtering by diffuse Ly\(\alpha\) would not significantly increase that estimate. For refractory solids, the FUV yields would be expected to be much lower still.



\paragraph{Sputtering from particle impact}

In general, the gas temperature in an ionized nebula is a factor of a few lower than the color temperature of the radiation field. As a result, the sputtering yield from particle impacts is expected to be much lower than the EUV photosputtering yield. Calculations of ion sputtering yields from silicates and carbon show that the yield only becomes appreciable for incident particle energies \(> 50\)~eV \citep{Field:1997}. In principle, this might be compensated by a very high impinging particle flux at the solid surface. However, taking thermal ion speeds of \(\sim 10~\mathrm{km\ s^{-1}}\) and densities of \(\sim 10^4~\mathrm{cm^{-3}}\), it is straightforward to show that the impinging particle flux is generally less than the EUV photon flux of equation~(\ref{eq:maximal-ablated-flux}). Therefore, particle sputtering is likely to be unimportant. 


\paragraph{Thermal sublimation}
\label{sec:thermal-sublimation}
Unlike the other processes considered so far, sublimation is a collective thermodynamic process that depends on the temperature of the solid body rather than directly on the properties of the impinging radiation. The radiative equilibrium temperature of a solid body at a distance \(10^{17} R_{17}~\mathrm{cm}\) from a star of bolometric luminosity \(10^4 L_4~L_\odot\) is
\begin{equation}
  \label{eq:radiative-equilibrium}
  T_{\mathrm{solid}} = 48.1\, L_4^{1/4} R_{17}^{-1/2} (Q_{\mathrm{abs}}/Q_{\mathrm{em}})^{1/4} \ \mathrm{K} ,
\end{equation}
where \(Q_{\mathrm{abs}}\) and \(Q_{\mathrm{em}}\) are respectively the absorption and emission efficiencies of the solid body with respect to a black body. The results of \S~\ref{sec:minim-size-ablat} rule out bodies smaller than meter-size, in which case \(a \gg \lambda\) and one has \(Q_{\mathrm{abs}} \sim Q_{\mathrm{em}} \sim 1\). The sublimation flux from the surface of water ice is such a strong function of temperature \citep{2007A&A...475..755G} that it may be considered a step function at \(T_{\mathrm{sub}} \simeq 150~\mathrm{K}\). From equation~(\ref{eq:radiative-equilibrium}), this may be transformed into a radius in the nebula within which sublimation will dominate photosputtering: \(R_{17} < 0.1 (150/T_{\mathrm{sub}})^2 L_4^{1/2}\). The bolometric luminosity of the central star will initially be rather high (\(L_4 \sim 1\)), falling to \(L_4 \sim 0.01\) once envelope hydrogen burning ceases \citep{1995A&A...299..755B}, with the higher-luminosity phase lasting for most of the lifetime of the nebula if the central star mass \(< 0.6~M_\odot\), but only a fraction of this time for higher mass central stars. Therefore, thermal sublimation of ices can be important locally during the early, compact phases of PN evolution (the first few hundred years, assuming an expansion velocity of \(20~\mathrm{km\ s^{-1}}\)), but proceeds at a rate much lower than maximally efficient photosputtering over the vast majority of the nebular lifetime and volume. Refractory materials, such as silicates, have a sublimation temperature that is roughly 10 times higher \citep[][e.g.,]{1989ApJ...345..230G} and so sublimation of these is never important beyond a few AU from the star.
 
\smallskip
In summary, none of these ablation mechanisms is likely to significantly reduce the ablation lifetime from the value calculated assuming maximally efficient photosputtering (equation~(\ref{eq:maximal-ablated-flux})). 

\subsection{Particle ablation velocity, \(v\ab\)}
\label{sec:part-ablat-veloc}

An upper limit to the velocity of the ablated metal ion can be found by assuming that its kinetic energy, \(E\ab\),  is a significant fraction of the incident photon energy, giving \(E\ab < 30\)~eV\@. The velocity is then \(v\ab = (2 E\ab/m\ab)^{1/2}\). 

\section{Results}
\label{sec:appl-minim-size}

\subsection{Minimum size of ablating solid bodies }
\label{sec:minim-size-ablat}

Substituting the value for the maximal ablation rate into equation~(\ref{eq:a-over-l}), one finds a dependence on \(n\neb/F_*\), which is proportional to the inverse of the ionization parameter, calculated in \S~\ref{sec:contr-therm-evap}. Hence, using the result of \S~\ref{sec:part-ablat-veloc}, the size constraint becomes 
\begin{equation}
  \label{eq:size-condition}
  \frac{a}{\ell} > 
  \left( \frac{z\neb \beta\ab}{3 Y\ab U} \right)^{1/2} = 
  \New{0.0012} \left(\frac{\New{z_{-3}^2} T_4 R_{17}}{Y\ab^2 S_{47} A_{16}}\right)^{1/4}  ,
\end{equation}
where \(A_{16} = A\ab/16\) and \New{\(z_{-3} = z\neb / 10^{-3}\) (for reference, the solar oxygen abundance is \(z_{-3} = 0.49 \pm 0.06\), \citealp{Asplund:2009})}. Note that this result is rather insensitive to changes in the nebular parameters.


From \S~\ref{sec:part-inter-scale}, a representative value of the interaction scale is \(\ell = 10^6 ~\mathrm{cm}\), so it can be seen from equation~(\ref{eq:size-condition}) that only the ablation of bodies larger than \(\sim 20~\mathrm{m}\) can produce cool, metal-rich pockets of ionized gas. In particular, this local constraint rules out dust grains, but allows larger bodies such as comets or planets.\footnote{In principle, there is a second regime of small
  sizes that can also give \(z\ab(\ell) > z_0\) due to a very large
  number of evaporating bodies being present within the volume
  \(\ell^3\) (see footnote~\ref{fn:crowding}). However, it can be
  shown that this condition is \( {a}/{\ell} < 3 \times 10^{-18} {f
    S_{47} A_{16}^{1/2}}/{(\rho\sol R_{17}^2)} , \) where \(f\) is the
  ratio of the total mass in solid bodies to the total nebular ionized
  gas mass and \(\rho\sol\) is the density of each solid body (in
  \(\mathrm{g\ cm^{-3}}\)). It can be easily seen that this size
  regime is unphysically small for any reasonable combination of
  parameters.}


\subsection{Minimum mass of solid body reservoir }
\label{sec:minimum-mass-solid}

Turning now to the global constraint of \S\S~\ref{sec:minimum-oxygen-mass} and \ref{sec:reservoir-mass}, the nebular mass can be estimated as 
\begin{equation}
  \label{eq:neb-mass}
  M\neb \simeq \frac43 \pi R\neb^3 n\neb m_{\mathrm{H}}
  \simeq 0.03 (S_{47} T_4 R_{17}^3)^{1/2} M_\odot . 
\end{equation}
so that equation~(\ref{eq:mass-condition}) for the minimum mass of ablated metals becomes
\begin{equation}
  \label{eq:mass-limit}
   M\ab \ge \New{9.0 \times 10^{-7} z_{-3}} \mu_{-3} 
   (S_{47} T_4 R_{17}^3)^{1/2} ~M_\odot .
\end{equation}

Using the an ablation rate of \(Y\ab\) times the maximal value of equation~(\ref{eq:maximal-ablated-flux}), we find an ablation lifetime of
\begin{equation}
  t\ab = 1.7 \times 10^8
  \left(\frac{a_5 \rho\sol R_{17}^2}{ A_{16} Y\ab S_{47}}\right) 
  ~\mathrm{years} , 
  \label{eq:ablate-lifetime}
\end{equation}
whereas the nebular age can be expressed in terms of its radius and a mean expansion velocity, which we normalize by a typical PN value of 20~km~s\(^{-1}\): 
\begin{equation}
  \label{eq:nebular-age}
  t = 2000 (R_{17} / v_{20})~\mathrm{years} .
\end{equation}
Hence, the minimum mass required in the solid body reservoir is given by 
\begin{equation}
  \label{eq:reservoir-mass-limit}
   M\sol \ge \New{0.045}
   \left(a_5 \rho\sol S_{47}^{-1/2} T_4^{1/2} R_{17}^{5/2} 
     \New{z_{-3}} \mu_{-3} v_{20} f\Ox^{-1} Y\ab^{-1}\right)
   ~M_\odot .
\end{equation}


\section{Discussion}
\label{sec:discussion}

\subsection{\New{Ongoing destruction of solids during the planetary nebula phase}}
\label{sec:ongo-destr-solids}

The constraints derived in the previous section \New{for destruction of solids during the planetary nebula phase} are very severe.  Solid bodies must be meter-sized or larger, ruling out classical interstellar or circumstellar dust grains. Furthermore, the total mass of solid bodies must exceed a few hundredths of a solar mass for kilometer-sized bodies, scaling linearly with the solid body radius. These limits apply to volatile materials, such as water ice, where the EUV photosputtering yield, \(Y\ab\),  may approach unity. Refractory materials, such as silicates, are likely to have a lower yield, which would require an even higher reservoir mass. Any dilution of the high metallicity gas by mixing with the background nebula over the lifetime of the PN would also only serve to tighten the mass constraint. Potential populations of solid bodies that might be found around the central star of a PN include planetary systems, comets, and debris disks. We will now examine each of these in turn. 

Planets can be ruled out as significant sources of gas phase metals on several grounds. First, they are expected to be found rather close to the central star, \(< 100\)~AU, which is much smaller than typical PN sizes, which are \( > 10^4\)~AU\@. Second, the most massive planets are likely to be gas giants, which, although they may be evaporated during the PN phase \citep{Villaver:2007}, are unlikely to contribute to oxygen enrichment. Rocky planets of size \(1000\)--\(10,000\)~km could potentially provide the metals, but equation~(\ref{eq:reservoir-mass-limit}) with \(a_5 = 10^4\) implies a minimum reservoir mass of \(\sim 1000~M_\odot\), which is obviously out of the question.

Comets in the Oort cloud around our sun have semi-major axes that typically lie in the range 1,000--100,000~AU, or 0.005--0.5~pc \citep{Morbidelli:2005}, which is similar to the sizes of planetary nebulae. The individual comets are mostly kilometer-sized, thus easily satisfying the local constraint of \S~\ref{sec:minim-size-ablat}. However, the total number of comets in our solar system has been variously estimated as \(10^{11}\)--\(10^{12}\) \citep{2007AA...461..741N}, with typical individual masses of \(4 \times 10^{16}~\mathrm{g}\) \citep{Heisler:1990}, giving a total mass of \(10^{-6}\)--\(10^{-5}~M_\odot\) for the comet reservoir. This is many times lower than the estimated minimum mass of \(\sim 0.03~M_\odot\) implied by equation~(\ref{eq:reservoir-mass-limit}). 

Debris disks around main sequence stars are detected via their FIR dust emission, but this dust is required to be constantly replenished by collisions among an unseen population of planetesimals \citep{Wyatt:2008}. The total mass of the debris disk is initially \NewNew{\(\sim 10^{-6}\)--\(10^{-4}~M_\sun\)}, and this is likely to have fallen by a factor a few by the time that the star reaches the AGB and PN phases of its evolution \citep{Lohne:2008}. Taking the minimum-sized body (\(\sim 10\)~m) that can satisfy equation~(\ref{eq:size-condition}), then equation~(\ref{eq:reservoir-mass-limit}) requires a reservoir mass of \(\sim 10^{-3}~M_\odot\). This already exceeds the total planetesimal mass \NewNew{of the most massive debris disks}, but the situation is even worse since models suggest that the debris disk mass is dominated by larger bodies, with the mass of meter-class rocks being as low as \(10^{-10}~M_\odot\) \citep[Fig.~5]{Lohne:2008}. \NewNew{Since the required reservoir mass scales linearly with the radius of the body, this would be \(\sim 0.1~M_\odot\) for kilometer sized bodies, which is again many times too large.} 

\subsection{\New{Comet evaporation during prior stellar evolutionary phases}}
\label{sec:comet-evap-during}

\New{The preceding sections have concentrated on the destruction of solid bodies during the relatively short-lived planetary nebula phase, but if a much longer span of time were available for the evaporation process, then a less massive solid body reservoir would be sufficient. The majority of the physical processes considered in \S~\ref{sec:maxim-ablat-rate} are inoperative during earlier stages of stellar evolution, since they depend directly or indirectly on the existence of ionizing photons. However, thermal sublimation (\S~\ref{sec:thermal-sublimation}) is an exception to this since it depends only on the bolometric luminosity of the central star, which is generally high (\(10^3\)--\(10^4~L_\sun\), e.g., \citealp{Marigo:2007}) for approximately \(10^6\) years while the star ascends the thermally pulsing asymptotic giant branch (AGB), immediately prior to the formation of the planetary nebula. Indeed, observations of water vapor emission in the circumstellar outflows of carbon-rich AGB stars \citep{2001Natur.412..160M} have been interpreted \citep{2001ApJ...557L.113S} as evidence for the sublimation during this evolutionary stage of a population of comets lying at a distance from the star of \(\simeq 100\)--\(200\)~AU, although current observations are insufficient to discriminate between this and competing mechanisms \citep{2007ApJ...669..412G}.  

The high-luminosity AGB phase typically lasts 100 to 1000 times longer than the planetary nebula phase \citep{1995A&A...299..755B, Marigo:2007}, so that the estimated minimum mass of the solid body reservoir (eq.~(\ref{eq:ablated-minimum-mass})) is reduced accordingly, to \(\sim 10^{-5}~M_\sun\), which is now comparable with estimates for the total mass of comets orbiting the sun (see above). A similar result was obtained from the more detailed comet sublimation models of \citet{2001ApJ...557L.113S}. However, even if mixing can be avoided over such a long timespan, it is not possible that comets evaporated during the entire AGB phase can explain the abundance discrepancies in planetary nebulae. This is because the AGB stellar wind speed (typically \(5\) to \(25~\mathrm{km\ s^{-1}}\), \citealp{1993A&AS...99..291L}) is a significant fraction of the subsequent expansion speed of the planetary nebula shell (\(20\)--\(30~\mathrm{km\ s^{-1}}\), \citealp{1989A&AS...78..301W, 2005ApJS..160..272H}), while the observed water vapor emission line has a similar width to CO lines from the same AGB star \citep{2007ApJ...669..412G}, implying that newly evaporated metals are efficiently entrained by the wind. Therefore, a planetary nebula of age \(t\) years will incorporate only those metals that were evaporated within a period of at most \(\simeq 6 t\) years before the onset of the planetary nebula phase. Any metals that were evaporated earlier than this, during the previous \(> 10^6\) years of AGB evolution \cite{Marigo:2007}, will lie in the low surface brightness outer halo of the planetary nebula \citep{1992ApJ...392..582B, 2002ApJ...571..880V, 2002ApJ...581.1204V} and so cannot contribute to observed abundance discrepancies. \NewNew{Since the bolometric luminosity of the star towards the end of the AGB phase (\(\sim 10^4~L_\odot\)) is very similar to that at the beginning of the planetary nebula  phase, the estimates given in \S~\ref{sec:thermal-sublimation} for the sublimation contribution would be increased by at most a factor of a few. 

To obtain a better quantitative estimate of this effect, we use the modeling results of \citet{2001ApJ...557L.113S}, who find an average water outflow rate of \(\simeq 10^{-6} M_\mathrm{solid}~\mathrm{yr}^{-1}\) during the late AGB phase, where \(M_\mathrm{solid}\) is the total ice mass in suitably situated bodies.\footnote{\NewNew{The peak rate during thermal pulses is \(\sim 10\) times higher, but the integrated contribution of these short bursts is small.}} The reciprocal of this quantity (\(10^6\) years) is equivalent to the ablation timescale, \(t_{\mathrm{ab}}\), of equation~(\ref{eq:ablate-lifetime}). Using equation (\ref{eq:nebular-age}) for the nebular age, together with the above estimate that a \(6\) times longer AGB mass-loss timespan can contribute material to the nebula, gives an effective timespan for sublimation of \(t \simeq 10^4\)~years. From equation~(\ref{eq:mass-limit}), the minimum evaporated oxygen mass for significant abundance discrepancies is \(\sim 10^{-6}~M_\odot\), thus the required comet reservoir mass is \(t\ab/t = 100\) times larger, or \(\sim 10^{-4}~M_\odot \simeq 33~M_\oplus\), for comets with astrocentric orbital radii in the narrow range \(200\)--\(300\)~AU\@. Closer-in comets are completely vaporised during the early AGB phase, whereas farther-out comets never attain high enough temperatures to undergo significant sublimation. The estimated mass of our Solar System's entire present-day Oort cloud \citep{2007AA...461..741N} is 10--100 times smaller than this, while the mass of the ``extended scattered disk'' \citep{Morbidelli:2005} that occupies the required range of orbital radii is much smaller still (\(\sim 0.01~M_\oplus\), \citealp{Gomes:2008}). However, the required comet reservoir mass is not much larger than the \(5\)--\(10~M_\oplus\) that is required\footnote{Assuming the same \(10^6\)~year ablation timescale as assumed above.} in order for comet evaporation to explain the H\(_2\)O \(1_{10}\)--\(1_{01}\) emission line from the carbon-rich AGB star IRC+10216 \citep{2001ApJ...557L.113S, 2007ApJ...669..412G}. 

Thus, evaporation of the same icy bodies could potentially explain both the presence of water vapor in a carbon-rich AGB stars and the abundance discrepancies that are observed in the subsequent planetary nebula phase, but only if the total volatile mass were orders of magnitude larger than any comet population observed in our Solar System. \citeauthor{2001ApJ...557L.113S} attempt to resolve this discrepancy by positing a much larger (\(\ge 10~M_\oplus\)) initial mass for the Kuiper belt, which was subsequently ejected to larger heliocentric radii, where it remains unobserved. However, more recent deep surveys of Trans-Neptunian objects have failed to find any evidence for such a population beyond 60~AU \citep{Fuentes:2008}. On the other hand, it is possible that the low mass of our present-day Kuiper belt is due to a delayed planetesimal-driven orbital migration of the giant planets \citep{Tsiganis:2005}, which destabilised the outer planetesimal disk, transferring many bodies into the inner solar system, where they collided with terrestrial planets \citep{Gomes:2005}, and others into the Oort cloud (3000--30,000~AU, \citealp{Brasser:2008}). Comparison with the correlation between age and mid-infrared excess in extrasolar debris disks \citep{Booth:2009} suggests that such disruptive scattering events are rare in other stellar systems and that many debris disks may retain a high mass (\(> 10~M_\oplus\)) through to the AGB phase. It should be noted, however, that only \(\simeq 16\%\) of A--K main-sequence stars show detectable debris disks \citep{Trilling:2008}, which renders problematic the application of this mechanism to all planetary nebulae. 

It is also worth pointing out that any mixing of the evaporated comet material with the AGB wind or with the photoionized nebula has the potential to decrease the efficiency of this mechanism. Mixing due to thermal diffusion can be calculated for both planetary nebula and AGB wind conditions by supposing that a comet of size \(\simeq 10\)~km is instantaneously evaporated and that the resultant high-metalicity cloud quickly reaches pressure and ionization balance with its surroundings. The cloud temperature is assumed to be \(100\)~K, which gives a cloud size of \(\sim 10^{11}\)~cm in both cases since the PN thermal pressure (\( P / k \sim 10^8~\mathrm{cm^{-3}\ K}\)) is very similar to the ram pressure in the AGB wind (assuming \(\dot M = 10^{-7}~M_\odot~\mathrm{yr}^{-1}\), \(V = 15~\mathrm{km\ s^{-1}}\), \(R = 200~\mathrm{AU}\)). Using mutual diffusion coefficients calculated as in \citet{Oey:2003}, one finds a thermal diffusion time for the cloud of \(\sim 10^9\)~years for the ionized planetary nebula case, but only \(\sim 10^4\)~years for the case of the neutral/molecular AGB wind. Diffusion is much more efficient in the AGB wind because the collisional mean free path is much longer due to the lack of long-range Coulomb interactions. This timescale is of the same order as the sublimation timescale calculated above, suggesting that diffusive mixing may be important in suppressing metallicity inhomogeneities. A more detailed study of this point, including turbulent mixing, will be the subject of future work. 
}

}

\section{Conclusions}
\label{sec:conclusions}

\NewNew{We have shown that solid body destruction \emph{during the planetary nebula phase itself} cannot explain observed abundance discrepancies. Although sufficiently massive populations of solid bodies may be present around planetary nebula central stars, the calculated sputtering rates are too low to ablate more than a small fraction of this mass during the lifetime of the planetary nebula. On the other hand, sublimation of volatile bodies during the final stages of the preceding asymptotic giant branch phase might possibly provide enough high-metallicity gas-phase material to explain the abundance discrepancies, but only if mixing is inefficient and in systems that possess a much more massive population of comets than is found in our Solar System. The inferred mass of solid bodies is similar to that required in order for comet evaporation to explain observations of water vapor in carbon-rich AGB stars, and is at the high end of the observed mass distribution of debris disks around solar-type stars.  }

\acknowledgments
WJH is grateful to LUTH, Observatoire de Paris for hospitality and
financial suport during two visits in which this work was carried
out. WJH also acknowledges financial support from DGAPA-UNAM, Mexico
(PAPIIT IN110108).  We thank Robin Williams for useful discussions and Peter van Hoof and Joe Weingartner for advice on photosputtering yields. \New{We are grateful to the anonymous referee for drawing our attention to the evidence for sublimation of comets during the AGB phase.}

\clearpage

\bibliographystyle{apj}
\bibliography{BibdeskLibrary}

\newcommand\Symbol[3]{%
\item[\(#1\)] #2 \(\mathrm{#3}\)
}


\newenvironment{SymbolList}{%
 \begingroup
   \footnotesize
   \renewcommand\baselinestretch{0.6}
    \begin{description}
    }{%
    \end{description}
 \endgroup
}

\begin{appendix}

\section{List of symbols}
\label{sec:list-symbols}
\twocolumngrid

\begin{SymbolList}
  \Symbol{\alpha\orl}{Effective recombination coefficient of O\(^{2+}\) line}{[cm^3\ s^{-1}]}
  \Symbol{\alpha_{\mathrm{B}}}{Case B hydrogen recombination coefficient}{[cm^3\ s^{-1}]}
  \Symbol{\beta\ab}{Dimensionless velocity of ablated particle: \(v\ab/c\)}{}
  \Symbol{\beta}{Plasma parameter: \(P_{\mathrm{gas}}/P_{\mathrm{mag}}\)}{}
  \Symbol{\mu_{\mathrm{min}}}{Minimum value of \(M_{\mathrm{O,1}} / M_{\mathrm{O,2}}\) that gives \(\mathrm{ADF} > 1\)}{}
  \Symbol{\rho\sol}{Bulk mass density of solid body}{[g\ cm^{-3}]}
  \Symbol{\sigma_{\mathrm{d}}}{Dust absorption cross-section}{[cm^2\ H^{-1}]}
  \Symbol{\tau}{Optical depth to ionizing radiation}{}
  \Symbol{a}{Radius of solid body}{[cm]}
  \Symbol{\mathrm{ADF}}{Abundance discrepancy factor: \(( \mathrm{O^{2+}}\!/ \mathrm{H^+} )_\mathrm{ORL} \,\smash{\big/}\, ( \mathrm{O^{2+}}\!/ \mathrm{H^+} )_\mathrm{CEL}\)}{}
  \Symbol{A\ab}{Atomic mass of ablated particle: \(m\ab/m_{\mathrm{H}}\)}{}
  \Symbol{E\ab}{Kinetic energy of ablated particle}{[erg]}
  \Symbol{f_{\mathrm{O}}}{Oxygen mass fraction of solid body}{}
  \Symbol{F\ab}{Ablative particle flux}{[atom\ cm^{-2}\ s^{-1}]}
  \Symbol{F_\alpha}{Flux of Lyman \(\alpha\) photons}{[cm^{-2}\ s^{-1}]}
  \Symbol{F_*}{Ionizing photon flux}{[cm^{-2}\ s^{-1}]}
  \Symbol{i}{Label for normal (\(i = i\)) and metal-enriched (\(i = 2\)) nebular gas phases}{}
  \Symbol{\ell}{Characteristic length scale for interparticle interactions}{[cm]}
  \Symbol{L_i}{Emission line luminosity of phase \(i\)}{[photon\ s^{-1}]}
  \Symbol{L_*}{Bolometric stellar luminosity}{[erg\ s^{-1}]}
  \Symbol{m\ab}{Mass of an ablated particle}{[g]}
  \Symbol{m\sol}{Mass of an individual solid body}{[g]}
  \Symbol{m_{\mathrm{H}}}{Mass of a hydrogen atom}{[g]}
  \Symbol{m}{Mass of atom or molecule}{[g]}
  \Symbol{M\ab}{Total mass of ablated metals}{[g]}
  \Symbol{M\neb}{Total mass of the ionized nebula}{[g]}
  \Symbol{M\sol}{Total mass of solid body reservoir}{[g]}
  \Symbol{M_{\mathrm{O},i}}{Total oxygen mass of phase \(i\)}{[g]}
  \Symbol{M_{\odot}}{Solar mass}{[g]}
  \Symbol{n\ab}{Number density of ablated metal atoms}{[cm^{-3}]}
  \Symbol{n_i}{Hydrogen number density of phase \(i\)}{[cm^{-3}]}
  \Symbol{n\neb}{Hydrogen number density of nebula}{[cm^{-3}]}
  \Symbol{\New{S_*}}{Stellar ionizing photon luminosity}{[s^{-1}]}
  \Symbol{Q_{\mathrm{abs}}}{Solid body absorption efficiency}{}
  \Symbol{Q_{\mathrm{em}}}{Solid body emission efficiency}{}
  \Symbol{r}{Distance of ablated particle from solid body}{[cm]}
  \Symbol{R}{Distance from star, radius of nebula}{[cm]}
  \Symbol{t\ab}{Timescale for destruction of solid body by ablation}{[s]}
  \Symbol{t}{Age of nebula}{[years]}
  \Symbol{T}{Gas temperature of nebula}{[K]}
  \Symbol{T_i}{Gas temperature of phase \(i\)}{[K]}
  \Symbol{T\sol}{Temperature of solid body}{[K]}
  \Symbol{T_{\mathrm{sub}}}{Sublimation temperature}{[K]}
  \Symbol{U}{Ionization parameter}{}
  \Symbol{v\ab}{Velocity of ablated particle}{[cm\ s^{-1}]}
  \Symbol{v}{Expansion velocity of nebula}{[cm\ s^{-1}]}
  \Symbol{V_i}{Volume of phase \(i\)}{[cm^{-3}]}
  \Symbol{y_i}{Ionization fraction, \(\mathrm{O^{2+}/O}\), in phase \(i\)}{}
  \Symbol{Y\ab}{Ablative yield}{}
  \Symbol{z}{Oxygen abundance by number relative to hydrogen}{}
  \Symbol{z_i}{Oxygen abundance of phase \(i\)}{}
  \Symbol{z\ab}{Oxygen abundance of ablated material (before mixing)}{}
\end{SymbolList}

\end{appendix}

\end{document}